# Current-induced giant diamagnetism in the Mott insulator $Ca_2RuO_4$

Chanchal Sow[1†], Shingo Yonezawa[1], Sota Kitamura[2],
Takashi Oka[3,4], Kazuhiko Kuroki[5],
Fumihiko Nakamura[6] & Yoshiteru Maeno[1†]

[1]Department of Physics, Graduate School of Science, Kyoto University, Kyoto 606-8502, Japan

[2]Department of Physics, Graduate School of Science, University of Tokyo, Tokyo 113-0033, Japan

[3]Max Planck Institute for Chemical Physics of Solids, D-01187 Dresden, Germany

[4]Max Planck Institute for the Physics of Complex Systems, D-01187 Dresden, Germany

[5]Department of Physics, Graduate School of Science, Osaka University, Osaka 560-0043, Japan

[6]Department of Education and Creation Engineering, Kurume Institute of Technology, Fukuoka 830-0052, Japan

†e-mail: chanchal@scphys.kyoto-u.ac.jp,    maeno@scphys.kyoto-u.ac.jp

*June 25, 2016*

**One-sentence summary:**

The strongest diamagnetism among non-superconducting materials emerges in a Mott insulator when it is tuned to semimetal by current.

**Mott insulators have surprised us many times by hosting new and diverse quantum phenomena when the "frozen" electrons are perturbed by various stimuli. Superconductivity, metal-insulator transition, and colossal magnetoresistance induced by element substitution, pressure, and magnetic field are prominent examples. Here we report a novel phenomenon, namely *giant diamagnetism*, in the Mott insulator $Ca_2RuO_4$ induced by electric current. With application of 1 $A/cm^2$ current, the strongest diamagnetism among all nonsuperconducting materials is induced as the system is tuned to a semimetallic state. The origin lies in the emergence of indirect Dirac cones in the many-body spectrum and associated monopole-like anomaly in the momentum dependent susceptibility. This record-breaking and switchable diamagnetism is a new class of non-equilibrium quantum phenomena on the verge of Mott insulating states.**

When one applies magnetic field to the conduction electrons in a material, they exhibit cyclotron motion resulting in orbital magnetic moment. Such orbital motion leads to quantized (discrete) energy levels known as the Landau levels. In a simple system, the resultant increase in the total energy is proportional to the square of the applied magnetic field. This increase in the energy results in negative magnetic susceptibility, which does not depend on the field strength. This effect is known as the Landau diamagnetism or orbital diamagnetism (*1*). Another source of diamagnetism, present ubiquitously irrespective of metalicity of a material, is circulation of inner-core electrons. In most materials, both types of diamagnetism are rather weak and often hindered by other paramagnetic contributions. However, pyrolytic graphite and bismuth are well known to exhibit exceptionally large diamagnetism, as shown in Fig. 1, due to strong Landau diamagnetism originating from light-mass electrons as well as multi-orbital effects. Indeed, these materials host *gapped* Dirac cones in the electronic dispersions with strong hybridization between the electron and hole bands (*2*). In such systems, interband effects induced by magnetic fields can lead to strong dimagnetism (*3, 4*). Recently, relatively large diamagnetism is observed in some topological semimetals such as TaAs, ascribable to the Weyl electrons in the bulk electronic state (*5*), again with Dirac-cone dispersion. We also note that superconductors show much stronger diamagnetism but only up to the critical magnetic fields.

In this Report, we show that the Mott insulator $Ca_2RuO_4$ under electric current exhibits the strongest diamagnetism among all known non-superconducting materials as presented in Fig. 1. Remarkably, even including superconducting materials, the diamagnetism of $Ca_2RuO_4$ at 7 T is comparable to that of $YBa_2Cu_3O_{7-\delta}$, which remains strongly diamagnetic even in such high fields owing to its unusually high superconducting transition temperature. As illustrated in Fig.1B-D, a truly novel feature is the switchability of the diamagnetism. From the point of view of device applications, our finding is attractive because we can control not only the Mott gap but also the large diamagnetism electronically using current. We note that negative magnetization due to local magnetic moments is found in certain ferrimagnets such as $YVO_3$, which consists of two or more magnetic sublattices (*6*). However, such negative magnetization is not categorized as diamagnetism by definition.

The Mott insulator $Ca_2RuO_4$ (*7*) is an end member of the system $Ca_{2-x}Sr_xRuO_4$, which exhibits a rich variety of magnetic, transport, and structural properties (*8–12*), including the spin-triplet superconductivity in $Sr_2RuO_4$ as the other end member (*13,14*). $Ca_2RuO_4$ exhibits metal to insulator transition at $T_{MI}$ = 357 K accompanied by the first-order structural transition with flattening of the $RuO_6$ octahedra below $T_{MI}$ (*15–17*). On further cooling, $Ca_2RuO_4$ undergoes an antiferromagnetic (AFM) ordering at $T_N$ = 113 K with the Ru spins aligning along the orthorhombic *b* axis as illustrated in Fig. 2A. Resonant x-ray diffraction reveals another order, known as orbital ordering (OO), at $T_{OO} \approx$ 260 K, below which the fourth electron of the $Ru^{4+}(4d^4)$ ion occupies the $d_{xy}$ orbital predominantly (*18*). $Ca_2RuO_4$ can be made metallic by pressure (*11*), chemical substitution (*8*), or electric field (*19*), accompanied by a first-order structural transition to stretch the RuO$_6$ octahedra along the *c*-direction. However, diamagnetism in the normal state has never been reported in Ca$_{2-x}$Sr$_x$RuO$_4$.

The phase diagram of Ca$_2$RuO$_4$ under electric current presented in Fig. 2C summarizes the present results. The original Mott insulating state without current evolves into a semimetallic state above a few mA (corresponding to the current density of 1 A/cm$^2$) and concurrently giant diamagnetism emerges at temperatures below 40 K. The current used in this Report is much lower than the current required to induce the insulator to metal transition at room temperature (5 A/cm$^2$) (*19*). Figure 2D represents the temperature dependence of the resistivity $\rho(T)$ under various applied currents for a sample with the cross-section of 3.2 × 10$^{-3}$ cm$^2$. Typical insulating behavior in $\rho(T)$ is observed at 10 nA. It is technically difficult to pass moderate current (e.g. 10 nA–3 mA) down to low temperatures due to high resistivity of the material (Fig. S1). However, with increasing current, it becomes possible to flow current down to 20 K because the resistivity drops by more than 5 orders of magnitude with current of as small as 4 mA. The shape of the $\rho(T)$ curve changes from the thermal-activation type to semimetallic behavior with increasing current, indicating partial closing of the Mott gap (*20*).

The magnetization $M(T)$ also exhibits substantial change by application of current. As shown in Fig. 2E, $M(T)$ under 10 nA is identical to the behavior at zero current (*7*). However, with increasing current, $M$ decreases gradually and the AFM transition vanishes at 3 mA. Current forces electrons to be itinerant and melts the AFM ordering. With higher current (410 mA), $M(T)$ exhibits a sharp drop below 50 K and continues to become negative at lower temperatures. We carefully checked that this negative signal is not an experimental artifact (Fig. S2). We find that the direction of the magnetization is always antiparallel to the applied field, regardless of the field and/or current directions (Fig. S3). Thus the local magnetic field induced by external current does not play a role. This invariance also excludes magnetoelectric effects observed in multiferroic systems (*21*) as a possible origin. We also emphasize that Joule heating, which certainly rises the actual sample temperature (*22*), cannot explain the diamagnetism, because $Ca_2RuO_4$ under ambient condition exhibits positive magnetic susceptibility for all temperatures. Therefore, we conclude that the observed negative magnetization in $Ca_2RuO_4$ originates from diamagnetism of conduction carriers due to current-induced partial Mott-gap closing.

Next, we focus on magnetotransport properties. As shown in Fig. 2F, the longitudinal magnetoresistance (MR) under 5 mA exhibits sign reversal at around 70 K. Such negative MR without apparent ferromagnetic spin fluctuation is unusual. We note that negative MR has been observed in some Weyl semimetals and attributed to the chiral anomaly, i.e. imbalance of *rightwinded* and *leftwinded* particles (*23*). Nevertheless, the origin of the negative MR in $Ca_2RuO_4$ is not clear yet. The Hall coefficient under 5 mA (Fig. 2G) also exhibits sign change below 70 – 80 K indicating presence of multiple types of carriers in the current-induced semimetallic state. The positive Hall coefficient below 70 K suggests the dominance of hole-like carriers.

The observed diamagnetism exhibits a peculiar anisotropy with respect to the applied magnetic field direction. For comparison, Fig. 3A shows the anisotropy of $M(T)$ with zero current, reflecting the direction of the local magnetic moment ordered parallel to the orthorhombic *b* axis (the $b_o$ axis) in the AFM state (*7*). For $M(T)$ with field along the *c* axis, a broad peak appears at about 250 K, attributable to the OO. With 5-mA current (Fig. 3B), negative *M* emerges below 40 K for all field directions. Reflecting a qualitative change in the magnetism, the anisotropy under 5 mA between $H // a_o$ and $H // b_o$ is smaller than the anisotropy at $I = 0$. Nevertheless, we found that the current-induced diamagnetism is substantially small for $H // a_o + b_o$ (45 degrees from the orthorhombic *a* axis to the *b* axis). Figure 3C shows the $M(H)$ curves at 20 K with the current $I = 0$ and 5 mA. The $M(H)$ response for $I = 0$ mA is positive and close to linear for all field directions, while $M(H)$ for $I = 5$ mA is negative and linear only up to 2-3 T; it becomes weakly sub-linear at higher fields. Figure 3D compares the resistivity normalized by the values at 300 K for *I* k *b* and for various field directions. This normalized resistivity is largest for *H // a*$_o$ + *b*$_o$, again manifesting a characteristic in-plane anisotropy.

The observed diamagnetic semimetal state is theoretically explained as a consequence of momentum-dependent Mott-gap closing due to hybridization of one-dimensional (1D) correlated orbitals. Here, we use a tight-binding approach and start with the band structure of a hypothetical metallic state for Ca$_2$RuO$_4$ without electron correlations. We assume the orbital ordered states: two electrons per Ru$^{4+}$ ion fully occupy the $d_{xy}$ orbitals and the other two partially occupy the $d_{yz}$ and $d_{xz}$ orbitals. Thus the electronic states close to the Fermi level consist of two half-filled bands originating from the $d_{yz}$ and $d_{xz}$ orbitals. These bands have one-dimensional nature with dispersions in the $k_x$ and $k_y$-directions. The two bands hybridize to form two quasi-two-dimensional (Q2D) cylindrical Fermi surfaces called α (hole band) and β (electron band) (Fig. 4A). The resulting Fermi surfaces are essentially what are observed in metallic Ca$_{1.8}$Sr$_{0.2}$RuO$_4$ (*24*), but different from that in metallic Ca$_2$RuO$_4$ above $T_{MI}$ where the structural transition makes all three bands form Fermi surfaces.

In order to account for the electronic correlation and the Mott insulating nature, we next employ the phenomenological approach proposed in Refs. (*25,26*). When the effect of correlation due to electron-electron interaction represented by the self energy Σ = Δ$^2$/(ω + $ε_{1D}(k)$) ($ε_{1D}(k)$: uncorrelated 1D dispersion) is turned on, a Mott gap Δ immediately opens up where the 1D nature is strongly preserved (on the Γ-M and M-X lines; Fig. 4B). In contrast, the Fermi surface remains ungapped near the Γ-X line, where the Q2D nature is strong. Such momentum-dependent gap opening leads to a Fermi-arc like structure. This Fermi arc shrinks as the electron correlation is increased, resulting in a pair of small electron-hole pockets on the Γ-X line (Fig. 4C). Although the wave numbers characterizing these pockets are slightly different (i.e. indirect gap), both pockets have "massive" Dirac-like dispersions (Fig. 4K). The formation of such indirect Dirac cones is owing to the hybridization of the underlining purely one-dimensional correlated bands. On further increasing the correlation, the gap fully opens completing the formation of the upper and lower Hubbard bands (Figs. 4I and 4J).

The indirect Dirac cones have important features common to those of the momentum-space Dirac monopoles (*27*). The inverse masses of the bands are light, while strong hybridization exists among them. In the case of Dirac monopoles, this leads to interesting Berry-curvature physics such as the anomalous Hall effect. Figure 4K is the three-dimensional plot of the correlated band energy as a function of $k_x$ and $k_y$ near the indirect Dirac cones for Δ = 24 meV. The associated momentum-dependent diamagnetic susceptibility Λ($k$) (*22*) indeed exhibits sharp peaks at these Dirac points (Fig. 4L). The total magnetic susceptibility is obtained by integrating Λ($k$) over the Brillouin zone (Fig. 4M). In the vicinity of the full gap opening (Δ = 24 meV), a large diamagnetism with a sharp drop in susceptibility is derived. We note that the sign change of the calculated magnetic susceptibility occurs at 32 K, agreeing reasonably with the experimental observation of the diamagnetic onset of $T$ = 40 K.

One of the remaining issues is the mechanism of the Mott gap closing by a DC electric current, although experimentally the occurrence of the gap closing has been established (*20*). There exist theoretical efforts to explain suppression of ferromagnetic ordering (*28*) and melting of charge density waves (*29*) in highly correlated electron systems under current. In $Ca_2RuO_4$, strong coupling to the structural phase transition as well as the Dirac-cone dispersion near the Mott gap most likely play essential roles in the smallness in the critical electric field (*19*).

This work demonstrated that the Mott insulator can be driven to a new electronic state under non-equilibrium steady condition induced by a simple DC electric current. Many other Mott insulators may exhibit novel electronic states by tuning their gaps by DC current. Our work demonstrates that DC current is a new and powerful tuning parameter to explore phases emerging around the Mott insulating state.

## Acknowledgements

We acknowledge discussions with J. Georg Bednorz. We also acknowledge technical support from M.P. Jimenez-Segura. This work was supported by JSPS KAKENHI No. JP26247060, as well as Nos. JP15H05851, JP15H05852, and JP15K21717 (Topological Materials Science).

## List of supplementary materials

Materials and Methods Figures S1-S6.

## References and Notes

1. L. Landau, *Zeitschrift fu¨r Physik* **64**, 629 (1930).

2. Z. Zhu, A. Collaudin, B. Fauque, W. Kang, K. Behnia, *Nature Physics* **8**, 89 (2012).

3. H. Fukuyama, *Prog. Theor. Phys.* **45**, 704 (1971).

4. Y. Fuseya, M. Ogata, H. Fukuyama, *J. Phys. Soc. Jpn.* **84**, 012001 (2014).

5. Y. Liu *et al.*, *J. Magn. Magn. Mater.* **408**, 73 (2016).

6. Y. Ren *et al.*, *Nature* **396**, 441 (1998).

7. S. Nakatsuji, S.-i. Ikeda, Y. Maeno, *J. Phys. Soc. Jpn.* **66**, 1868 (1997).

8. S. Nakatsuji, Y. Maeno, *Phys. Rev. Lett.* **84**, 2666 (2000).

9. O. Friedt *et al.*, *Phys. Rev. B* **63**, 174432 (2001).

10. T. Mizokawa *et al.*, *Phys. Rev. Lett.* **87**, 077202 (2001).

11. F. Nakamura *et al.*, *Phys. Rev. B* **65**, 220402 (2002).

12. S. Nakatsuji *et al.*, *Phys. Rev. Lett.* **93**, 146401 (2004).

13. Y. Maeno *et al.*, *Nature* **372**, 532 (1994).

14. K. Ishida *et al.*, *Nature* **396**, 658 (1998).

15. C. S. Alexander *et al.*, *Phys. Rev. B* **60**, R8422 (1999).

16. 16. E. Gorelov *et al.*, *Phys. Rev. Lett.* **104**, 226401 (2010).

17. M. Braden, G. Andre, S. Nakatsuji, Y. Maeno, *Phys. Rev. B* **58**, 847 (1998).

18. I. Zegkinoglou *et al.*, *Phys. Rev. Lett.* **95**, 136401 (2005).

19. F. Nakamura *et al.*, *Sci. Rep.* **3** (2013).

20. R. Okazaki *et al.*, *J. Phys. Soc. Jpn.* **82**, 103702 (2013).

21. W. Eerenstein, N. Mathur, J. F. Scott, *Nature* **442**, 759 (2006).

22. Materials and Methods are available as Supplementary Materials on Science website.

23. X. Huang *et al.*, *Phys. Rev. X* **5**, 031023 (2015).

24. M. Neupane *et al.*, *Phys.Rev. Lett.* **103**, 097001 (2009).

25. R. M. Konik, T. M. Rice, A. M. Tsvelik, *Phys. Rev. Lett.* **96**, 086407 (2006).

26. K.-Y. Yang, T. Rice, F.-C. Zhang, *Phys. Rev. B* **73**, 174501 (2006).

27. Z. Fang *et al.*, *Science* **302**, 92 (2003).

28. A. Mitra, A. J. Millis, *Phys. Rev. B* **77**, 220404 (2008).

29. S. Ajisaka, H. Nishimura, S. Tasaki, I. Terasaki, *Progr. Theoret. Phys.* **121**, 1289 (2009).

30. D. Shoenberg, M. Z. Uddin, *Proc. R. Soc. Lond. A Math. Phys. Sci.* **156**, 701 (1936).

31. A. V. Narlikar, *Studies of high temperature superconductors: advances in research and applications*, vol. 14 (Nova Science Publishers, 1995).

32. F. Nakamura *et al.*, *Phys. Rev. B* **80**, 193103 (2009).

33. P. L. Alireza *et al.*, *J. Phys.: Condens. Matter* **22**, 052202 (2010).

34. F. Sawano *et al.*, *J. Phys. Soc. Jpn.* **78**, 024714 (2009).

35. I. Terasaki, R. Okazaki, H. Ohta, *Scr. Mater.* **111**, 23 (2016).

36. H. Fukuyama, R. Kubo, *J. Phys. Soc. Jpn.* **27**, 604 (1969).

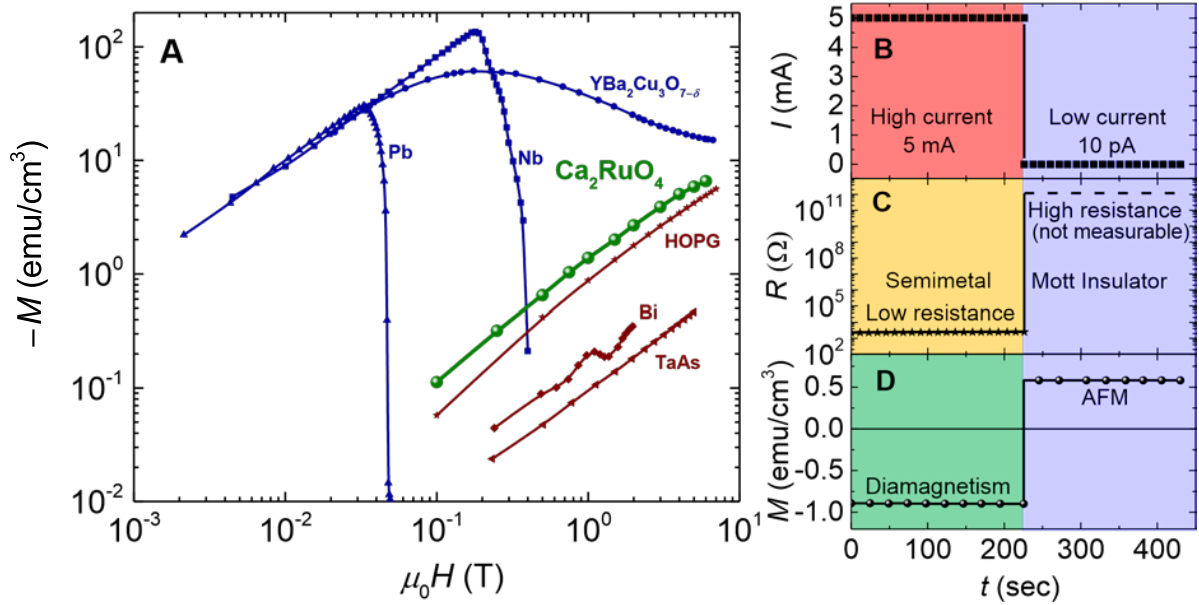

Fig. 1: **Giant and switchable diamagnetism observed in Ca$_2$RuO$_4$. A,** Comparison of current-induced diamagnetism in Ca$_2$RuO$_4$ with other common diamagnets and superconductors. Here we plot $-M$ as a function of applied magnetic field for Ca$_2$RuO$_4$ at 5 mA (1.5 A/cm$^2$) and 20 K, highly ordered pyrolytic graphite (HOPG) at 20 K, bismuth at 14 K (*30*), TaAs at 5 K (*5*), Pb (type-I superconductor) at 4.7 K (*31*), Nb (type-II superconductor) at 4.7 K (*31*), and YBa$_2$Cu$_3$O$_{7-\delta}$ (type-II superconductor) at 4.2 K (*31*). The data for Ca$_2$RuO$_4$ and HOPG were obtained by ourselves, whereas the rest of the *M*(*H*) data are collected from the cited articles. **B-D,** Experimental demonstration of switching of the giant diamagnetism using current. The resistance (panel **C**) and magnetization (panel **D**) in Ca$_2$RuO$_4$ drastically changes as the current (panel **B**) was turned off. This experiment was performed at *T*=20 K and $\mu_0 H$=1 T. The resistance at 10 pA was much higher than 10$^{11}$ Ω, the upper limit of resistance measurable with our setup.

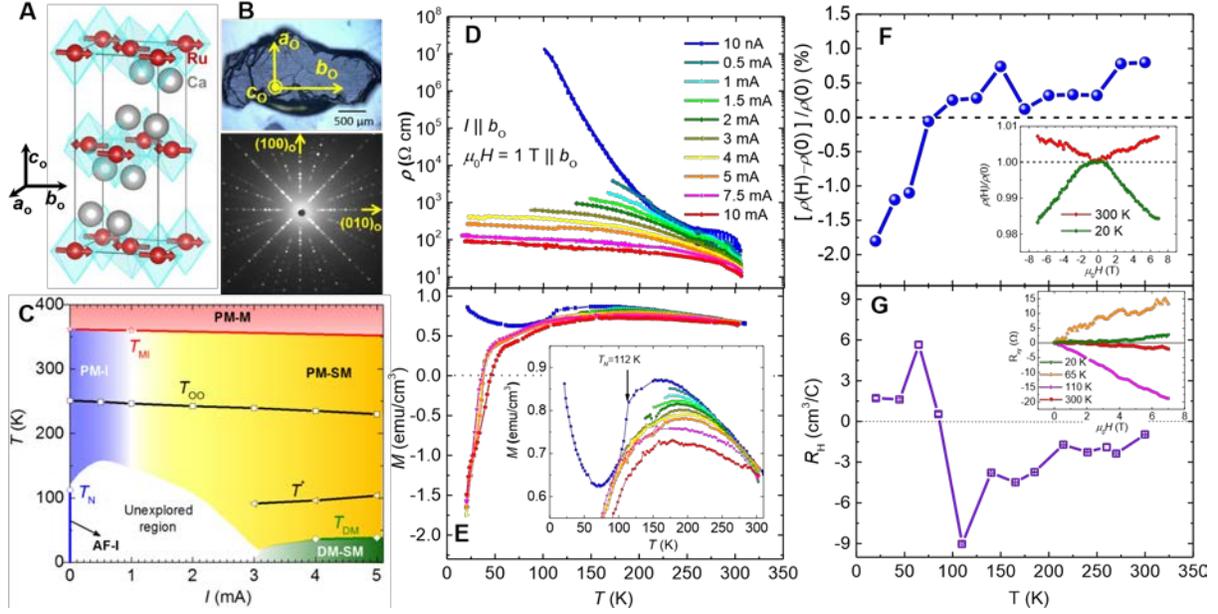

Fig. 2: **Novel diamagnetic semimetal state in $Ca_2RuO_4$ induced by current. A,** Crystal structure of $Ca_2RuO_4$. The primitive vectors $a_o$, $b_o$, and $c_o$, are defined in the orthorhombic notation. The magnetic structure in the AFM state under zero current is illustrated by arrows. **B,** Optical microscope image of a single-crystal sample with a Laue pattern. **C,** Phase diagram of $Ca_2RuO_4$ under current revealed in this study. 5-mA current corresponds to the current density of 1.5 A/cm$^2$. We have used abbreviations PM-M (paramagnetic-metal), PM-I (paramagnetic insulator), PM-SM (paramagnetic-semimetal), AF-I (antiferromagnetic-insulator), and DM-SM (diamagnetic-semimetal). **D,** Temperature dependence of the resistivity for various applied current. **E,** Temperature dependence of the magnetization for various applied current. For the data in **D,** and **E,** current and the magnetic field are applied along the $b_o$ axis. **F,** The temperature dependence of the magnetoresistance ratio $[\rho(H) - \rho(0)]/\rho(0)$. The inset shows $\rho(H)/\rho(0)$ as a function of applied magnetic field at 20 and 300 K. **G,** Hall coefficient as a function of temperature. The inset shows the Hall resistance as a function of magnetic field measured at various temperatures.

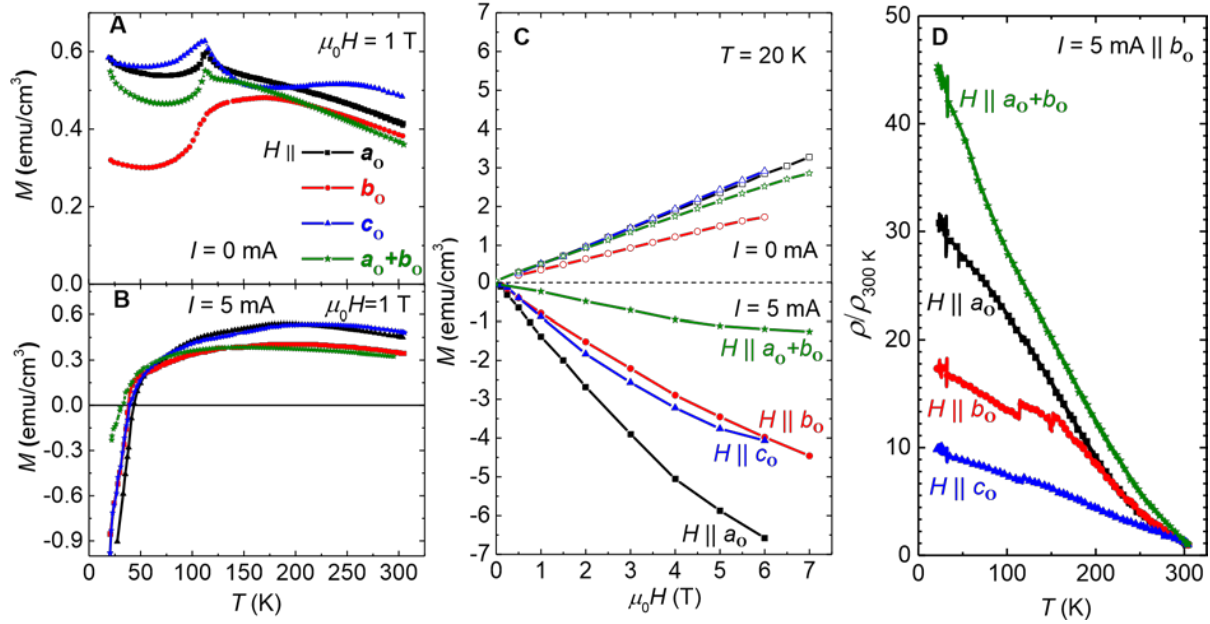

Fig. 3: **Magnetic-field anisotropy of the diamagnetic semimetallic behaviour in $Ca_2RuO_4$.** Anisotropy of the magnetization as a function of temperature with **A,** zero applied current and **B,** 5-mA current. **C,** Magnetic-field dependence of the magnetization for $I$ = 0 mA and 5 mA. **D,** Normalized resistivity $\rho(T)/\rho(300\,K)$ under 1-T field applied in various directions. The current was applied parallel to the $b_o$ axis for all cases.

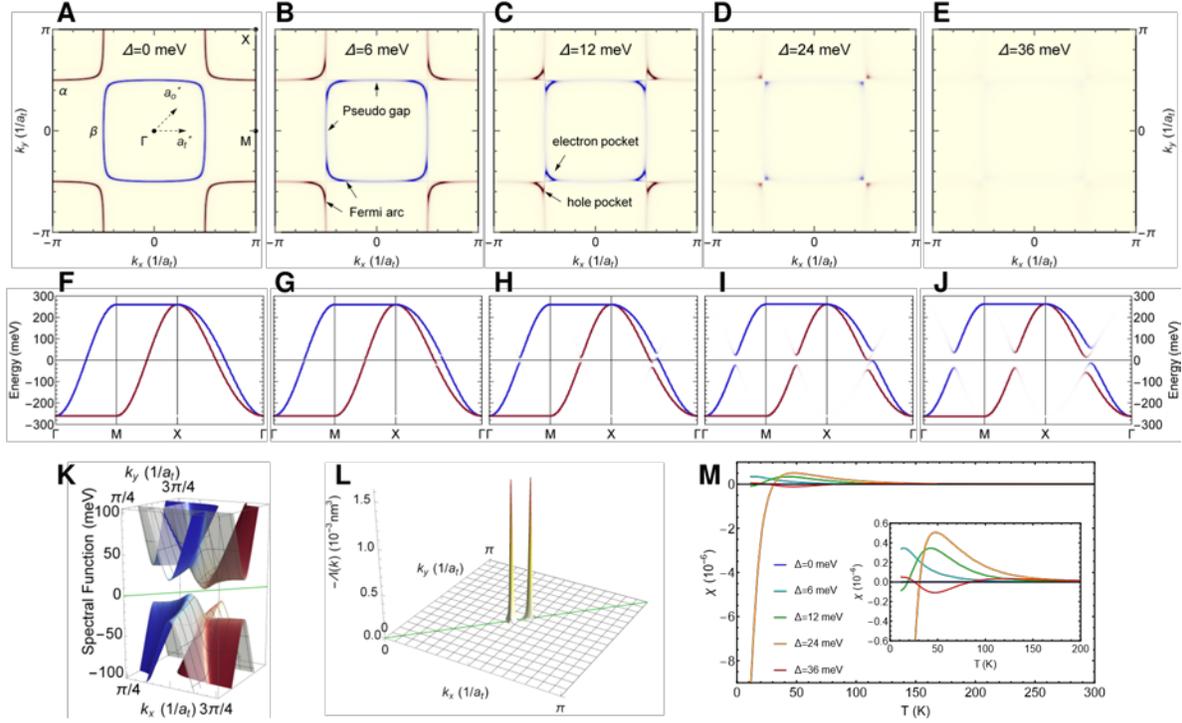

Fig. 4: **Electronic structure and associated diamagnetism of the indirect Dirac semimetal state of $Ca_2RuO_4$ induced by electric current. A-E,** Two-dimensional projection of the Fermi surface at various bare Mott gap Δ, starting from the tight-binding model for $Ca_2RuO_4$. **F-J,** Correlated band (zero points of the Green function) structures corresponding to the cases **A-E**. Blue and red lines indicate the β (hole-like) and α (electron-like) bands and the brightness of the color corresponds to the spectral weight at each point. **K,** Three-dimensional plot of the correlated band. **L,** Momentum-resolved diamagnetic susceptibility Λ(**k**) (at T =12 K) for Δ = 24 meV (case **D**). **M,** Calculated total magnetic susceptibility as a function of temperature. Large diamagnetic susceptibility emerges when the tips of the indirect Dirac cones touches the chemical potential (Δ = 24 meV).
14

Supplementary Materials for

# Current-induced giant diamagnetism in the Mott insulator Ca$_2$RuO$_4$

Chanchal Sow[1†], Shingo Yonezawa[1], Sota Kitamura[2],
Takashi Oka[3,4], Kazuhiko Kuroki[5],
Fumihiko Nakamura[6] & Yoshiteru Maeno[1†]

[1]Department of Physics, Graduate School of Science,
Kyoto University, Kyoto 606-8502, Japan

[2]Department of Physics, Graduate School of Science,
University of Tokyo, Tokyo 113-0033, Japan

[3]Max Planck Institute for Chemical Physics of Solids, D-01187 Dresden, Germany

[4]Max Planck Institute for the Physics of Complex Systems,
D-01187 Dresden, Germany

[5]Department of Physics, Graduate School of Science,
Osaka University, Osaka 560-0043, Japan

[6]Department of Education and Creation Engineering,
Kurume Institute of Technology, Fukuoka 830-0052, Japan

†e-mail: chanchal@scphys.kyoto-u.ac.jp, maeno@scphys.kyoto-u.ac.jp

*June 25, 2016*

This PDF file includes:

Materials and Methods

Figs. S1 to S6 with figure captions

## Materials and Methods

**Sample preparation and characterizations.** Single crystals of $Ca_2RuO_4$ were grown by a floating-zone method with $RuO_2$ self-flux using a commercial infrared furnace (Canon Machinery, model SC-M15HD). To obtain a high-density feed rod, mixtures of $CaCO_3$ (99.99%) and $RuO_2$ (99.9%) were calcined at 1000°C in air for 10 h. The crystal growth was performed at a rate of 10 mm/h inside the quartz tube of the infrared furnace. The obtained single-crystalline rod has a typical dimension of 4 mm in diameter and 50 mm in length. However, the rod breaks into pieces when cooled across the metal-to-insulator transition temperature (357 K). Nevertheless, we obtain high-quality single crystals with a typical dimension of 2-3 × 1 × 0.2 $mm^3$. We have confirmed by x-ray diffraction that the crystals have the so called S-Pbca structure (*7*) at room temperature. In this study, we used high quality single-crystalline samples with essentially stoichiometric oxygen content judging from the lattice parameter *c* of 11.92 Å at 290 K and the antiferromagnetic transition temperature of 113 K (*11,32,33*).

**Measurement set-up.** We designed a special sample holder made of non-magnetic (weakly diamagnetic) glass epoxy, connected to a long and thin stainless-steal tube. The holder fits inside the commercial SQUID magnetometer (Quantum Design, MPMS-XL) to measure the resistance and magnetization simultaneously under current. We put electrical leads inside the sample-holder tube. We have also placed a thermometer close to the sample. In order to get the magnetic signal only from sample we placed the temperature sensor 2 cm away from the sample but on the same copper strip as the sample was mounted. We have used a temperature monitor (Lakeshore, 218) to read the temperature.

The transport measurements were carried out with a combination of an electrometer (Keithley, 6514) and current Source (Keithley, 6221) using the standard two-probe technique. The output impedance of the electrometer is higher than 200 TΩ. However, because of the resistance between the circuit and the ground of our sample holder, the upper limit of the measurable resistance was around $10^{11}\,\Omega$. We made gold contact pads using dc sputtering with an aluminum mask on top of crystals and used room-temperature-cure silver paint to ensure good contacts through thin copper wires. Although we used the two-probe technique, we confirmed that the sample resistance (no less than 180 Ω at room temperature) is much larger than other remnant resistances (resistance of wires, contact resistance, etc.) which is measured to be less than 1 Ω. Thus the measured sample resistance is still accurate.

It is known that DC current can cause considerable amount of Joule heating, especially for insulating materials. In order to dissipate the heat efficiently: (i) the sample is mounted on a long copper plate which is fixed firmly with the sample holder; (ii) most of the top surface of the sample is covered with silver paint which is in good thermal contact (using copper wire and GE varnish) to the holder. The sample is cooled directly with cold helium gas which allows to cool the sample from all sides. We have measured the actual temperature of the sample under 5 mA and 0 mA current. At 20 K, we get a difference of 3.5 K.

**Calculation methods.** Our theoretical calculation for the many-body band is based on the theory developed by Konik, Rice, and Tsvelik (*25*). First, we describe the system as a pair of one-dimensional Mott insulators ($d_{xz}$ and $d_{yz}$), whose Green function is given by

$$G_{1D}(\omega, k)^{-1} = \omega + 2t \cos k - \frac{\Delta^2}{\omega - 2t \cos k}, \tag{S1}$$

where $t$ denotes the nearest-neighbor hopping energy within the orbital and Δ is the Mott gap for the 1D system. The effect of correlation due to electron-electron interaction is represented by the self energy $\Sigma = \Delta^2/(\omega + \varepsilon_{1D}(k))$, where $\varepsilon_{1D}(k)$ is the bare dispersion of the 1D orbital. The two orbitals resides on a square lattice and we take into account the hybridization between the two orbitals within the random phase approximation, i.e.,

$$G(\omega, \boldsymbol{k})^{-1} = \begin{pmatrix} G_{1D}(\omega, k_x)^{-1} & -4g \sin k_x \sin k_y \\ -4g \sin k_x \sin k_y & G_{1D}(\omega, k_y)^{-1} \end{pmatrix}, \quad (S2)$$

where $g$ denotes the interorbital hopping between the next-nearest sites. Following experimental observations (*20,34,35*), we take into account the effect of the electric current by decreasing the bare Δ. The other parameters in the Green function are estimated by performing the firstprinciples calculation, yielding $t$ = 130 meV and $g$ = 11 meV. These parameters are consistent with Gorelov *et al.* (*16*). Here the effect of band folding due to scattering with a momentum transfer (π,π,0) originating from the tilt and rotation of RuO$_6$ octahedra is neglected. We have confirmed that such lattice change does not significantly affect the energy dispersion in the Mott-insulating regime.

The diamagnetic susceptibility with the field along *z*-axis is calculated using the Fukuyama formula (*36*), $\chi = \sum_{\boldsymbol{k}} \Lambda(\boldsymbol{k})$, where

$$\Lambda(\boldsymbol{k}) = \frac{e^2}{c^2} T \sum_n \mathrm{Tr}[G(i\omega_n, \boldsymbol{k})v_x G(i\omega_n, \boldsymbol{k})v_y]^2 \quad (S3)$$

is the momentum-dependent susceptibility up to bubble diagrams. Both intraband and interband contributions are incorporated in this expression. Here, $\omega_n$ is the Matsubara frequency and $v_{x,y}$ is the velocity operator. Expression (S3) for the diamagnetic susceptibility has a striking resemblance with the Kubo formula for Hall conductivity,

$$\sigma_{xy} = \lim_{\omega \to 0} \frac{i}{\omega} \sum_{\boldsymbol{k}} [Q_{xy}(\boldsymbol{k}, \omega + i\delta) - Q_{xy}(\boldsymbol{k}, 0)] \text{ with}$$

$$Q_{xy}(\boldsymbol{k}, i\nu_m) = -eT \sum_n \mathrm{Tr}\,[G(i\omega_n, \boldsymbol{k})v_x G(i\omega_n - i\nu_m, \boldsymbol{k})v_y]. \quad (S4)$$

The momentum dependent susceptibility $Q_{xy}(\boldsymbol{k},iv_m)$ is related to the Berry curvature in the zero temperature limit and its Brillouin zone integral gives a topological index, i.e., the Chern number. It is known that the Berry curvature shows a large peak near the Dirac monopole in momentum space (27). In a similar way, we find that the diamagnetic susceptibility $\Lambda(\boldsymbol{k})$ has a large peak at the position of the indirect Dirac cones (Fig. 5M). This monopole-like anomaly, however, has a slightly different nature when compared to the Hall susceptibility or diamagnetism in Dirac band materials. Namely, in the present case, contribution from intraband terms is dominant in producing the large diamagnetic response.

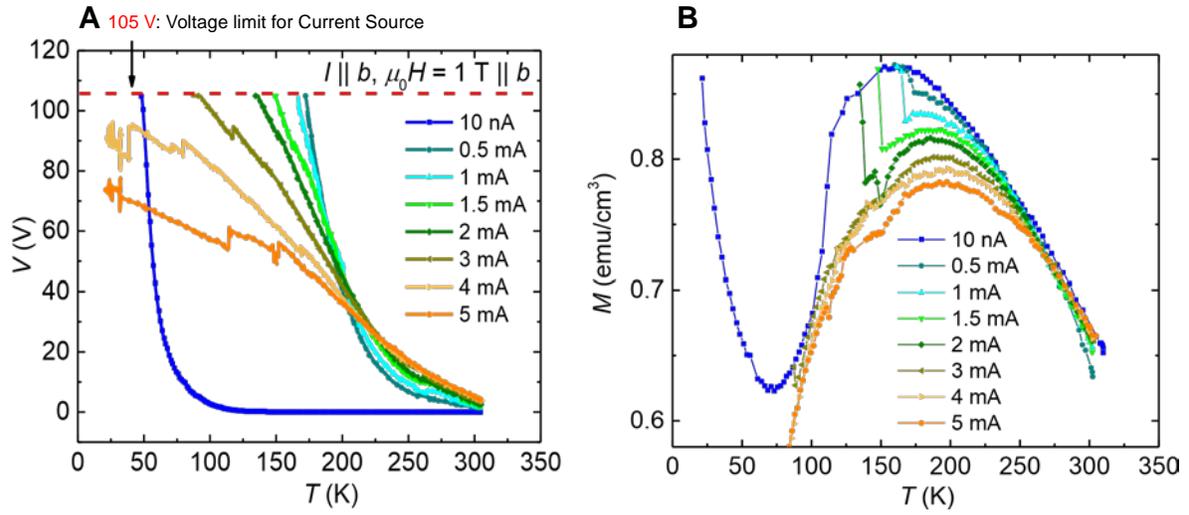

Fig. S1: **Technical details of the current application.** Simultaneous measurements of **A,** the voltage across the sample and **B,** magnetization, as a function of temperature at various applied current. In case of 10 nA - 3 mA, the current source stopped passing current when the voltage exceeds 105 V, which is the upper voltage limit of our current source (Keithley, 6221). When such current stop occurs, the magnetization jumps to a point on the 10 nA curve. This fact indicates that the sample quality is not changed by current.

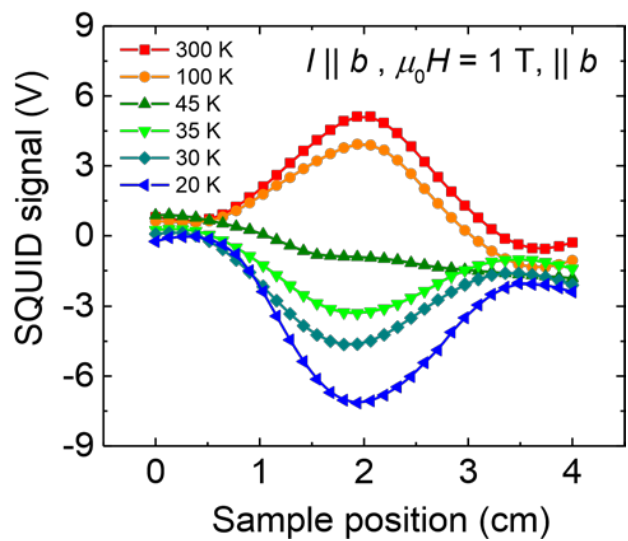

Fig. S2: **Raw SQUID signal of the magnetization measurement.** Representative SQUID signals during magnetization measurements of $Ca_2RuO_4$ under 5-mA current as a function of the sample positions for selected temperatures. The peak is located at 2 cm for all curves, indicating that the sample is symmetrically scanned in a pair of pick-up coils separated by 1.5 cm. The peak at 2 cm changes from positive to negative as temperature decreases, indicating a change from a paramagnetic state to a diamagnetic state. The small asymmetry of the signal originates from background contribution (the sample holder and thermometer).

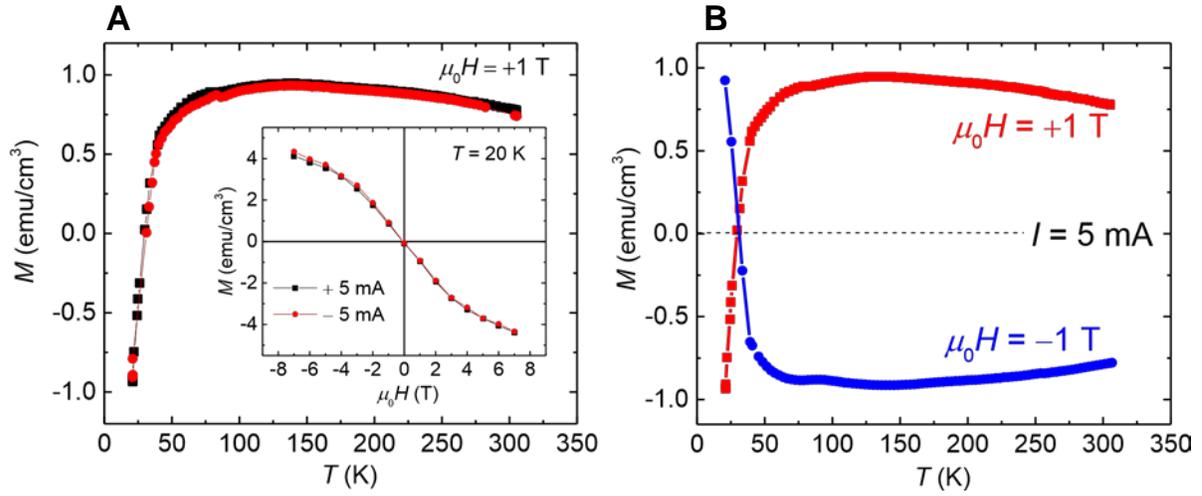

Fig. S3: **Field and current direction dependences of the magnetization.** The magnetization value of $Ca_2RuO_4$ under current below ~ 40 K is found to be negative, as explained in the main text. A naive question is whether the magnetic field induced by the applied current plays a role here or not. To answer this question, we investigate the field-direction and current direction dependences. **A,** $M(T)$ of $Ca_2RuO_4$ for $H // b$ measured under $I$ = +5 and −5 mA. The inset compares $M(H)$ at 20 K for $I$ = +5 and −5 mA. **B,** $M(T)$ under $I$ = +5 mA in $\mu_0 H$ = 1 T and −1 T applied parallel to the $b$ axis. These data clearly demonstrate that the induced magnetization remains antiparallel to the applied magnetic field, regardless of the current and field directions. This behavior is characteristic of the diamagnetism. If the current-induced magnetic field were the origin of the negative magnetization, the direction of the observed signal would be flipped when the current is reversed. Thus, it is clear that the current induced magnetic field is not the origin of the negative magnetization. The observation also rules out the possibility of the electric-field induced magnetic moment, known as the magnetoelectric effect observed in multiferroic materials.

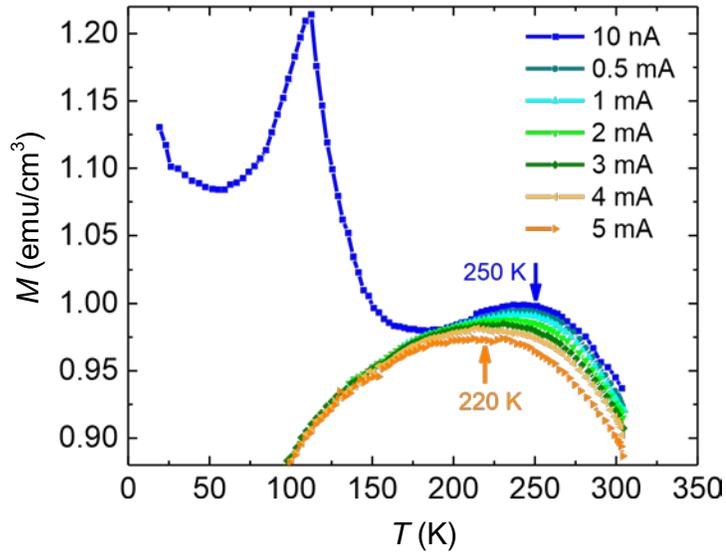

Fig. S4: **Shift of the orbital order by external current.** The magnetization of $Ca_2RuO_4$ for $H // c$ exhibits a broad hump located at 250 K, associated with the orbital ordering (*18*). The temperature of this hump decreases gradually with the increase in current. This peak temperature, $T_{OO}$ is plotted in the phase diagram shown in Fig. 2C (main text). Such gradual change of $T_{OO}$ with the application of current implies that the current-induced Mott-gap closing occurs within the same orbital states, in accordance with our theoretical model.

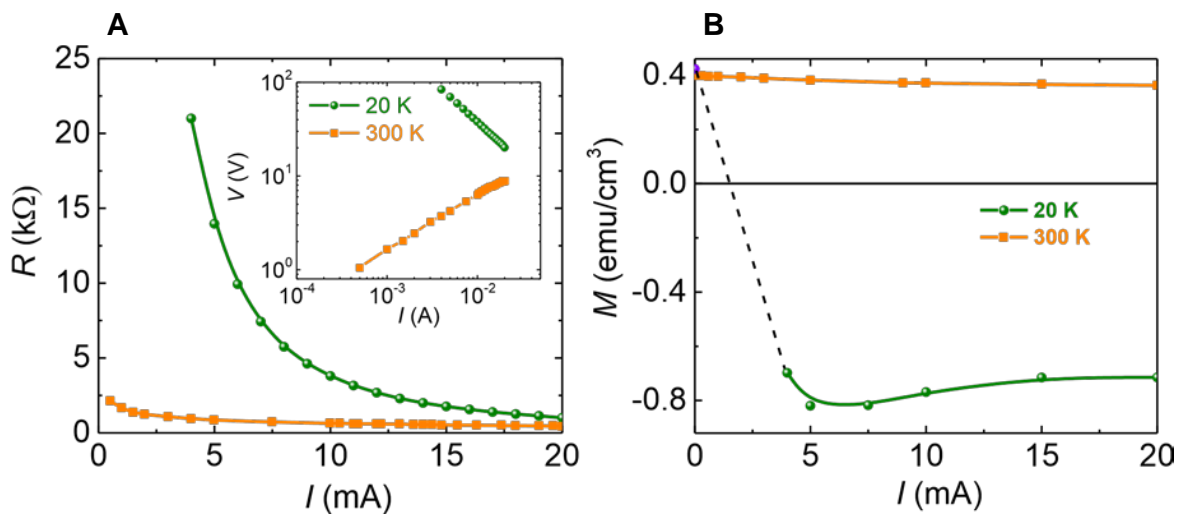

Fig. S5: **Current strength dependence of transport and magnetic properties. A,** Resistance as a function of applied current along the $b_o$ axis at 20 K (green circles) and 300 K (orange squares). The current dependence of the resistance at 20 K exhibits a power-law decrease with increasing $I$. This decrease indicates that the number of carrier substantially increases with increasing $I$. The inset shows the $I - V$ characteristics in log scales. At 20 K, the curve shows a negative slope (negative differential resistance, NDR), while at 300 K the slope is positive. **B,** Magnetization as a function of current at 20 K and 300 K. At 300 K, the magnetization is reduced gradually with increasing current but there is no signature of diamagnetism. However, diamagnetism appears at 20 K under 4-mA current. We note that the region between 10 nA and 4 mA was not experimentally accessible due to limitation of the current source as described above. Interestingly, within the diamagnetic state, the size of the magnetization does not change much with the increase in current above 4 mA.

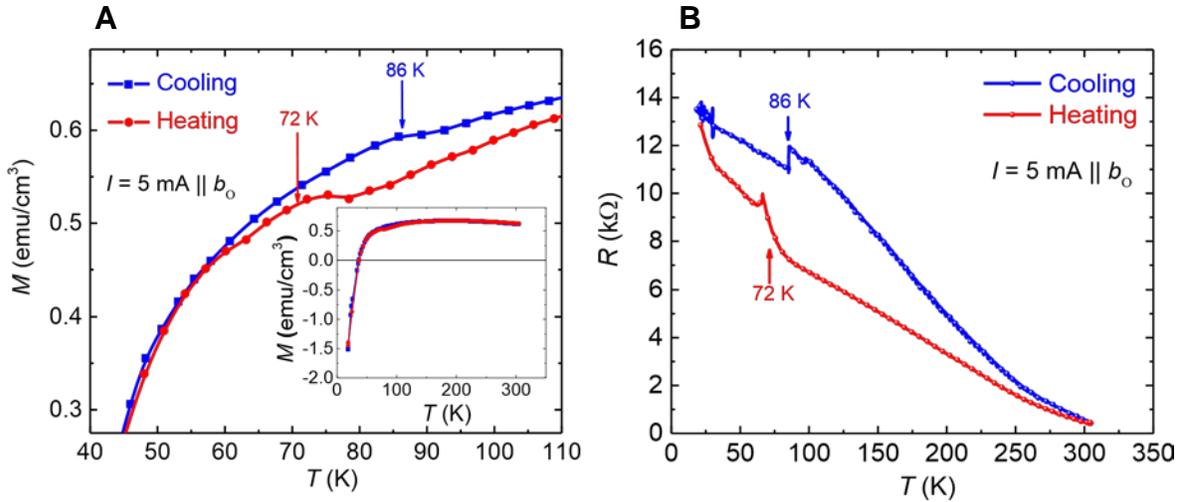

Fig. S6: **Possible current-induced first-order transition.** A careful inspection of $M(T)$ under current (*e.g.* Fig. 2E and Fig. S3) reveals an anomalous kink at ~ 80 K. In addition, small jump in $R(T)$ was also observed at similar temperatures. To obtain further insight, simultaneous measurements of **A,** $M(T)$ and **B,** $R(T)$ was performed during cooling and heating processes under 5-mA current and $\mu_0 H = 1$ T for $H // I // b_o$. The hump in $M(T)$ and a jump in $R(T)$ occur at nearly the same temperature within a process. We also found that the changes in $M(T)$ and $R(T)$ are accompanied by thermal hysteresis: the anomaly occurs at ~ 70 K in the warming process. Such hysteretic behavior indicates a first-order phase transition at around 80 K. However, the nature of this transition is not yet known.